\documentclass[twocolumn,aps,prd,amsmath,amssymb,nofootinbib,nobibnotes,floatfix,superscriptaddress]{revtex4-1}

\usepackage{graphicx}
\usepackage{float}
\usepackage{bm}
\usepackage[hidelinks]{hyperref}
\usepackage{color}

\usepackage[utf8]{inputenc}
\usepackage[T1]{fontenc}
\usepackage{slashed}

\usepackage{enumerate}

\usepackage{mathalfa,amssymb,amsthm}
\usepackage{amsmath}

\newcommand{\ssec}[1]{\section{#1}}

\newcommand{\beq}{\begin{equation}}
\newcommand{\eeq}{\end{equation}}
\newcommand{\beqa}{\begin{eqnarray}}
\newcommand{\eeqa}{\end{eqnarray}}

\begin{document}

\title{
    Instability and backreaction of massive spin-2 fields around black holes
   }
\author{William E.\ East}
\email{weast@perimeterinstitute.ca}
\affiliation{Perimeter Institute for Theoretical Physics, Waterloo, Ontario N2L 2Y5, Canada}
\author{Nils Siemonsen}
\email[]{nsiemonsen@perimeterinstitute.ca}
\affiliation{Perimeter Institute for Theoretical Physics, Waterloo, Ontario N2L 2Y5, Canada}
\affiliation{Arthur B. McDonald Canadian Astroparticle Physics Research Institute, 64 Bader Lane, Queen's University, Kingston, Ontario K7L 3N6, Canada}
\affiliation{Department of Physics and Astronomy, University of Waterloo, Waterloo, Ontario N2L 3G1, Canada}

\begin{abstract}
    A massive spin-2 field can grow unstably around a black hole, giving rise to a potential
    probe of the existence of such fields.
    In this work, we use time-domain evolutions to study such instabilities.
    Considering the linear regime by solving the equations generically
    governing a massive tensor field on the background of a Kerr black hole, we
    find that black hole spin increases the growth rate and, most
    significantly, the mass range of the axisymmetric (azimuthal number $m=0$) instability, which
    takes the form of the Gregory-Laflamme black string instability for zero
    spin.  We also consider the superradiant unstable modes with $1 \leq m \leq
    3$, extending previous results to higher spin-2 masses, black hole spins,
    and azimuthal numbers.  We find that the superradiant modes grow slower
    than the $m=0$ modes, except for a narrow range of high spins and masses,
    with $m=1$ and 2 requiring a dimensionless black hole spin of $a_{\rm
    BH}\gtrsim 0.95$ to be dominant. Thus, in most of the parameter space, the
    backreaction of the $m=0$ instability must be taken into account when using
    black holes to constrain massive spin-2 fields. As a simple
    model of this, we consider nonlinear evolutions in quadratic gravity, in
    particular Einstein-Weyl gravity. We find that, depending on the initial
    perturbation, the black hole may approach zero mass with the curvature
    blowing up in finite time, or can saturate at a larger mass with a
    surrounding cloud of the ghost spin-2 field. 
\end{abstract}
\maketitle
\ssec{Introduction}%
Recently, there has been a renewed interest in studying the phenomenological
implications of massive spin-2 particles, including as dark matter
candidates~\cite{Babichev:2016bxi,Aoki:2014cla,Marzola:2017lbt,GonzalezAlbornoz:2017gbh,Alexander:2020gmv,Jain:2021pnk,Manita:2022tkl,Kolb:2023dzp}.
Massive spin-2 fields also arise in a number of modifications of general
relativity~\cite{Clifton:2011jh}, including when adding quadratic curvature
terms to the action~\cite{Stelle:1976gc,Stelle:1977ry}, string-theory
compactifications~\cite{Buchbinder:1999ar}, nonlinear massive
gravity~\cite{deRham:2010kj,deRham:2014zqa}, and ghost-free bigravity
theories~\cite{Hassan:2011hr,Hassan:2011zd,Schmidt-May:2015vnx}.  While
nonlinear massive gravity requires the Vainshtein mechanism to recover
Newtonian gravity, bigravity theories with a massive and massless graviton
naturally recover general relativity with a weakly coupled massive spin-2 field, and thus are
commonly used to construct ghost-free, nonlinear theories of spin-2 dark
matter.

A powerful way to probe the existence of ultralight bosons that may be weakly
coupled to standard model matter is through the superradiant instability of
black holes (BHs), which only relies on the fact that the bosons
gravitate.  The
spin-0~\cite{Ternov:1978gq,Zouros:1979iw,Detweiler:1980uk,Cardoso:2005vk,Dolan:2007mj,Arvanitaki:2009fg,Arvanitaki:2010sy,Yoshino:2013ofa,Arvanitaki:2014wva,Brito:2014wla,Yoshino:2015nsa}
and spin-1~\cite{Rosa:2011my,Pani:2012vp,Pani:2012bp,Cardoso:2018tly,Baryakhtar:2017ngi,East:2017mrj,East:2018glu,Baumann:2019eav,Siemonsen:2019ebd},
cases have been well studied, leading to a detailed picture of the
observational implications (see Ref.~\cite{Brito:2015oca} for a review). In the presence of a spinning BH, a cloud
of massive bosons will grow exponentially at the expense of the rotational
energy of the BH. The instability will be fastest when the
Compton wavelength of the boson is comparable to the size of the BH, or
equivalently, when $\alpha:=\mu M$ is order one, where $m_b=\mu \hbar$ is the
boson mass, $M$ is the total spacetime mass, and we use
geometric units with $G=c=1$ throughout.  In the absence of significant
nongravitational interactions~\cite{Fukuda:2019ewf,Baryakhtar:2020gao,East:2022ppo,Cannizzaro:2022xyw}, the boson cloud will grow until the BH 
has been spun down sufficiently so that its horizon frequency matches the
oscillation frequency of the cloud~\cite{East:2017ovw,East:2018glu,Brito:2014wla}. Superradiance can thus be observationally
probed by measuring BH spins~\cite{Arvanitaki:2009fg,Cardoso:2018tly,Brito:2017zvb,Baryakhtar:2017ngi,Ng:2020ruv}, as well as searching for gravitational
wave signals sourced by the clouds oscillations~\cite{Brito:2017wnc,Brito:2017zvb,Ghosh:2018gaw,Isi:2018pzk,Tsukada:2019,Tsukada:2020lgt,LIGOScientific:2021rnv,KAGRA:2022osp,Chan:2022dkt,Jones:2023fzz}.

Massive spin-2 fields are also subject to the BH superradiant
instability, with even shorter timescales. While the nonlinear
behavior of a spin-2 field is model dependent, the linear limit around a
background spacetime like a BH is universally described by the
covariant Fierz-Pauli theory, meaning such instabilities will be a generic
feature of a large class of theories~\cite{Buchbinder:1999ar,Mazuet:2018ysa}. The superradiant instability of spin-2
fields has been studied in the nonrelativistic limit ($\alpha \ll 1$) using
semianalytic methods in Refs.~\cite{Brito:2013wya,Brito:2020lup}, and recently
for the fastest growing dipolar mode (azimuthal number $m=1$ mode) for $\alpha
\leq 0.8$ and dimensionless BH spins up to $a_{\rm BH}=0.99$ in Ref.~\cite{Dias:2023ynv}.  However,
an additional complication is that massive spin-2 fields are unstable to a
monopolar ($m=0$) instability even around nonspinning (Schwarzschild) BHs
when $\alpha \lesssim 0.4$~\cite{Babichev:2013una,Brito:2013wya,Myung:2013doa,Lu:2017kzi,Held:2022abx}. Intriguingly, this linear instability
takes the same form as the Gregory-Laflamme instability of a black string~\cite{Gregory:1993vy}
when identifying $\mu$ with the wave number along the flat direction of the string,
and the competition between superradiant and Gregory-Laflamme instabilities has
been studied in six spacetime
dimensions in Refs.~\cite{Dias:2022mde,Dias:2022str,Dias:2023nbj}.
It is not known how the monopolar instability extends
to spinning BHs, nor what the saturation of this instability
is (see Ref.~\cite{Brito:2013xaa} for one possibility). Thus, in order to use BH instabilities to probe massive spin-2 fields, it is essential
to know which instability dominates in different parts of the parameter space and what the backreaction
of that instability is.  

In this work, we tackle these questions using time domain evolutions of spin-2
fields. Considering unstable modes for different BH spins, values of
$\alpha$, and azimuthal numbers, we find that the $m=0$ instability dominates
over the fastest growing ($1\leq m \leq 2$) superradiant
modes in most of the parameter space (including $\alpha \lesssim 1$ and $a_{\rm
BH}\lesssim 0.95$, and all the previously studied parameter space). We find
that the superradiant instability is only the fastest growing mode for higher
azimuthal numbers, where the growth timescale is parametrically longer, or at
high BH spins and a narrow range of BH masses for lower
azimuthal numbers. 

To begin to address the effect of the monopolar massive spin-2 instability on
the BH, we here consider nonlinear evolutions in the modified theory where one
adds quadratic curvature terms to the Einstein-Hilbert action, in particular
Einstein-Weyl theory~\cite{Stelle:1976gc,Stelle:1977ry}. This theory has ghost
degrees of freedom associated with having fourth order equations of motion, and
will in general have different behavior than a ghost-free nonlinear theory.
However, it has a well-posed nonlinear
formulation~\cite{Noakes:1983xd,Morales:2018imi,Held:2021pht}, and serves here
as a simple model to illustrate several features of the backreaction of the
instability on a BH. We find that when the initial massive tensor perturbation
has the opposite sign energy to the BH, the BH grows in mass until the instability
saturates with a surrounding ghost spin-2 field.  When the perturbation carries
positive energy, we find the BH mass approaches zero in finite time.

\ssec{Model and Methodology\label{sec:methodology}}%
The linear evolution of a tensor field $H_{ab}$ with mass parameter $\mu$ on a Ricci-flat background
spacetime is governed by~\cite{Buchbinder:1999ar,Brito:2013wya,Mazuet:2018ysa}
\begin{equation}
    \label{eqn:lin_eom}
    \Box H_{ab} = \mu^2 H_{ab}-2R_{acbd}H^{cd}, \ H^a_a=\nabla_a H^{ab}=0  
\end{equation}
where $\Box=\nabla_a\nabla^a$ is the covariant wave operator. 
In this work, we evolve these equations on a Kerr BH
background 
in order to
identify unstable modes. Extending the techniques of Ref.~\cite{East:2017mrj} from
spin-1 to spin-2, we will consider different values
of $\alpha:=M\mu$, $a_{\rm BH}$, and $m$ (where, in coordinates adapted
to the axisymmetric Killing vector, $H_{ab} \sim  e^{im \phi}$, and we
have introduced an imaginary component to keep track of the angular phase---see Appendix~\ref{app:eqns}
for details),
and in each case evolve some perturbation until 
it is dominated by the fastest growing mode with those parameters.
We are primarily interested in measuring the complex frequencies of the most
unstable modes, $H_{ab}\sim e^{-i \omega t }$ with $\omega = \omega_R+i \omega_I$, by determining
the growth rate and oscillation frequency.  
In particular, we monitor the evolution of the conserved (up to flux through the BH horizon) quantity 
\begin{equation}
E := -\int H^t_a t^a \sqrt{-g} d^3 x \ , 
\end{equation}
where $t^a$ is the Killing vector associated with the stationary spacetime and $g$ is the metric determinant, and 
perform linear fits to $\arg E(t)$ and $\log |E(t)|$ after $|E|$ has grown through several $e$-folds.
We find equivalent results measuring the growth of other quantities (e.g. $H_{ab}H^{ab}$).

One can construct a theory coupling a massive spin-2 field to gravity without ghosts, which gives a nonlinear
extension of Eq.~\eqref{eqn:lin_eom}, using ghost-free bigravity~\cite{deRham:2010kj,Hassan:2011hr,Hassan:2011zd}
(with metric perturbations governed by the linearized Einstein equations).
However, developing a well-posed dynamical formulation of such theories is still a work in progress~\cite{deRham:2023ngf}.
Here, instead, as a toy-model of the nonlinear evolution of the BH instabilities, we will temporarily set aside our fear of ghosts
and consider vacuum Einstein-Weyl gravity. This theory has the following action
\begin{equation}
    S= \int d^4x \sqrt{-g} \left ( R -\frac{1}{2\mu^2} C^{abcd}C_{abcd}\right)
\end{equation}
where $C^{abcd}$ is the Weyl tensor.  Solutions to vacuum Einstein-Weyl gravity
are special cases of solutions to quadratic gravity (also known as fourth order
gravity~\cite{Barth:1983hb,Whitt:1985ki} or Stelle
gravity~\cite{Stelle:1976gc,Stelle:1977ry}) where the massive scalar degree of
freedom corresponding to the Ricci scalar in the latter is not
excited\footnote{In quadratic gravity, the Ricci scalar obeys a massive wave
equation and would thus be susceptible to the scalar superradiant instability,
but on much longer timescales than those considered here.}. The evolution equations are given in terms of the
(trace-free) Ricci tensor as
\begin{equation}
    \label{eqn:nonlin_eom}
    \Box {R}_{ab} = \mu^2 {R}_{ab}-2R_{acbd}{R}^{cd} + \frac{1}{2}g_{ab} {R}^{cd} {R}_{cd} , \ R^a_a=0 \ .
\end{equation}
These equations involve fourth-order derivatives of the metric, and the massive tensor
field can carry positive or negative energy as dictated by Ostrogadsky's theorem~\cite{Woodard:2015zca},
but do give rise to a well-posed evolution system in a generalized harmonic formulation~\cite{Noakes:1983xd,Morales:2018imi}.
Linearizing around a Ricci-flat background, Eq.~\eqref{eqn:nonlin_eom} reduces
to Eq.~\eqref{eqn:lin_eom} with $R_{ab}$ identified with $H_{ab}$~\cite{Myung:2013doa,Dias:2023ynv}.

We consider axisymmetric evolutions of the nonlinear equations~\eqref{eqn:nonlin_eom} and track the BH apparent horizon 
and monitor its area $A_{\rm BH}$, angular momentum $J_{\rm BH}$, and, through the Christodoulou formula, compute an associated mass $M_{\rm BH}$.
More details on the evolution equations, horizon diagnostics, numerical scheme, and convergence results can be found
in Appendices~\ref{app:eqns} and~\ref{app:conv}.

\ssec{Linear instability results}
%
\begin{figure*}
\includegraphics[width=0.85\textwidth]{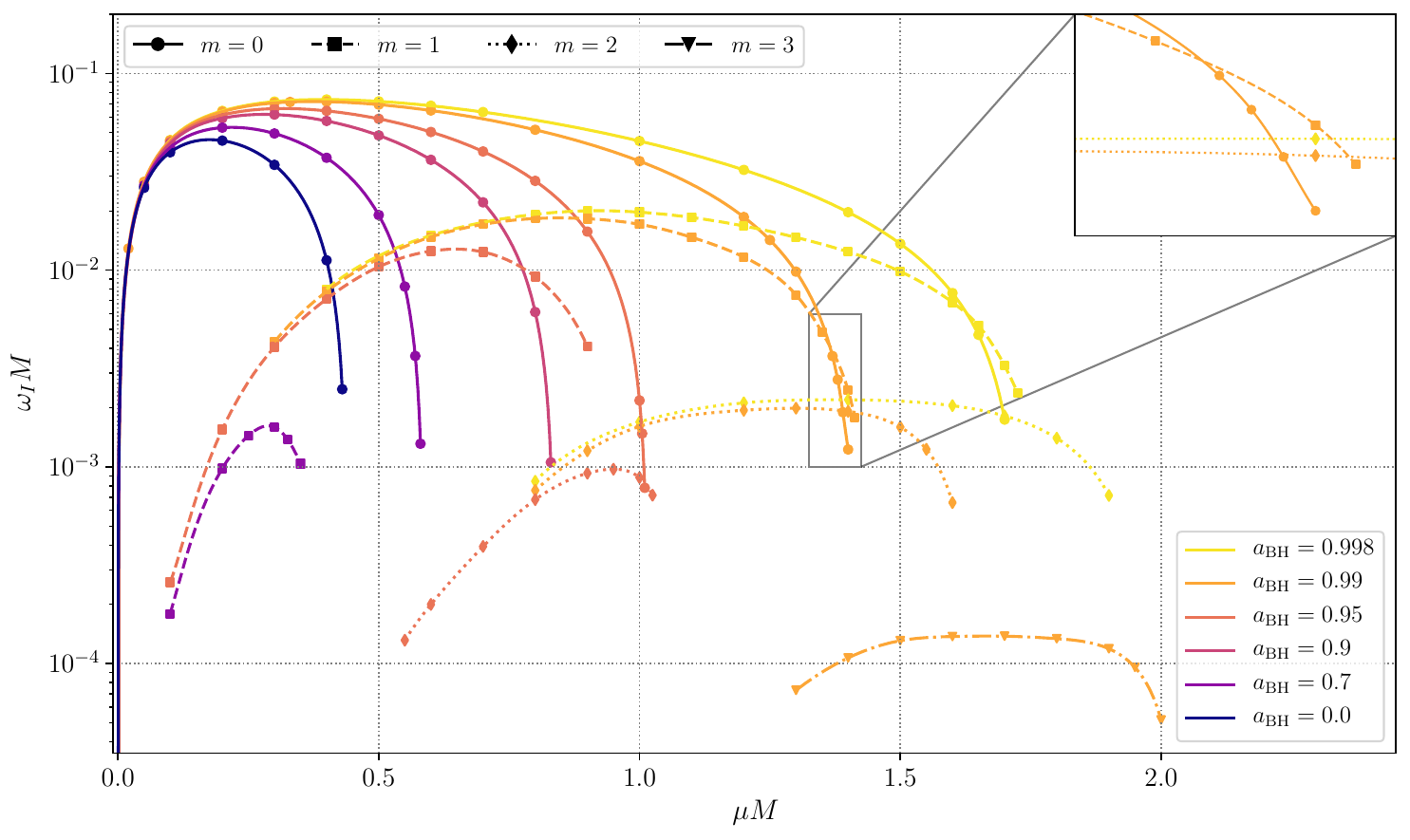}
    \caption{The growth rates $\omega_I$ of linear massive spin-2 field perturbations propagating in a Kerr background spacetime of mass $M$ and dimensionless spin $a_{\rm BH}$, characterized by their azimuthal mode number $m$, as a function of the mass parameter $\mu$. Data points indicate the rates extracted from our time-domain evolutions (available at Ref.~\cite{spin2_data}), while the lines connecting the points are cubic interpolations.}
\label{fig:growth rates}
\end{figure*}

We begin by studying the linear stability of massive
spin-2 perturbations on a Kerr BH background, which is generically
governed by Eq.~\eqref{eqn:lin_eom}.
In Refs.~\cite{Babichev:2013una,Brito:2013wya}, it was
noted that Schwarzschild BHs are unstable to a monopolar instability
($m=0$) with purely imaginary frequency: $\omega_I \sim \mu$. 
Kerr BHs are unstable to the growth of $m>0$ superradiant massive
spin-2 modes, as shown in Ref.~\cite{Brito:2013wya,Brito:2020lup,Dias:2023ynv}.
The fastest growing dipolar ($m=1$) mode was identified perturbatively in the
$\alpha\ll 1$ regime, with scaling $\omega_I M\sim \alpha^3$, while $\omega_I
M\sim \alpha^{2m+5}$ for $m\geq 2$ in this
regime. Hence, the monopolar instability should dominate in the $\alpha\ll
1$ limit. 
Therefore, a natural question arises: where in the parameter space is the
monopolar instability the fastest?

To address this question, we extend these analyses by considering (i) the monopolar instability on a Kerr background spacetime up to $a_{\rm BH}=0.998$, (ii) the most unstable dipolar superradiant family of modes in the highly relativistic limit of the parameter space, and (iii) higher-order unstable modes of the spin-2 field with $m=2$ and $m=3$. In Fig.~\ref{fig:growth rates}, we compare the growth rates $\omega_I$ of all these modes across the relevant parameter space, obtained using the methods outlined in Sec.~\ref{sec:methodology} (with details in Appendix~\ref{app:eqns}).

Focusing on the monopolar modes first, the growth rates follow the
spin-independent scaling $\omega_I\approx 0.62 \mu$ in the $\alpha\ll 1$
regime. These unstable modes turn stable at a critical mass $\mu_c$. This
critical point goes from $\mu_c M\approx 0.44$ in the Schwarzschild case (which is consistent with Ref.~\cite{Brito:2013wya}) up to
$\mu_c M\approx 1.73$ for $a_{\rm BH}=0.998$, near the extremal limit. The
maximum of the monopolar instability rate surpasses the rate of the slowest
decaying quasinormal mode of the BH when $a_{\rm BH}\gtrsim 0.91$. The
$m\geq 1$ superradiant modes exhibit the expected scaling in the $\alpha\ll 1$
limit~\cite{Brito:2020lup}, and show good agreement with the results in Ref.~\cite{Dias:2023ynv}
in the region of overlap. (See Appendix~\ref{app:modes} for a detailed comparison 
to Refs.~\cite{Brito:2020lup,Dias:2023ynv}, including the values of
$\omega_R$.) In the $\alpha\sim\mathcal{O}(1)$ regions of the parameter space,
these modes turn stable when the superradiance condition is saturated $\omega_R=m\Omega_H$
(where $\Omega_H$ is the BH horizon frequency).

Comparing the monopolar ($m=0$) mode with the most unstable superradiant $m=1$
mode, it becomes clear from Fig.~\ref{fig:growth rates} that the monopolar
instability dominates the dynamics of the system in the linear regime across
large ranges of the relevant parameter space. In fact, only in the near-extremal
limit, for critical spin of $a^c_{\rm BH}> 0.95$, are the growth rates of the
most unstable $m=1$ superradiant configuration comparable or larger than those
of the monopolar instability. (For comparison, we note that if instead one assumed
that the value of the critical mass for the monopolar mode remained roughly
constant with black hole spin at the Schwarzschild value of $\mu_c M\approx 0.44$, 
this would give a critical spin of $a^c_{\rm BH}\sim 0.72$ for the $m=1$ superradiant
instability to dominate~\cite{Dias:2023ynv}.)

Considering higher order superradiant modes, this
critical spin reduces with increasing azimuthal index $m$. From
Fig.~\ref{fig:growth rates}, $a^c_{\rm BH}<0.95$ for the $m=2$ superradiantly
unstable configurations and
$a^c_{\rm BH}\sim 0.7$ for $m=3$.%
\footnote{
Due to the longer growth timescales of the $m=3$ modes, we
were unable to confidently identify a growing mode at lower spins,
and this estimate is based on extrapolating
$\omega_R^{m=3}$ for $a_{\rm BH}=0.7$ to $3\Omega_H$.} 
Note, however, the maximum growth rate at
fixed spin decreases roughly exponentially with increasing azimuthal index. 

\ssec{Nonlinear evolution}
The results in the previous section indicate that the $m=0$ instability is the
fastest growing massive tensor mode around a Kerr BH for much of the
parameter space, leading naturally to the question:
what is the nonlinear development of this instability? 
Due to the connection with the Gregory-Laflamme instability~\cite{Gregory:1993vy} and questions of
cosmic censorship~\cite{Lehner:2010pn,Figueras:2022zkg}, this of theoretical, in addition to
phenomenological interest. The answer will in general depend
on the particular nonlinear model chosen, and will not be fully addressed here.
However, to gain some insight into the possibilities, we will consider nonlinear evolutions in
Einstein-Weyl gravity, restricting to axisymmetry. 

To begin with, we note a peculiarity of this theory is that, since the linearly
growing tensor field is just the (trace-free) Ricci tensor $R_{ab}$, if
we think of the backreaction on the BH metric as occurring through an
effective stress-energy tensor, in this case this would
just be proportional to $R_{ab}$ itself, and hence linearly (and not
quadratically) dependent on the exponentially growing mode. Hence, for any set
of parameters, we find two possible types of nonlinear behavior depending on the
initial sign of the perturbation: one corresponding to the BH mass
decreasing at the expense of the growing massive tensor field, and the other
corresponding to the BH mass growing.

\begin{figure}
\begin{center}
\includegraphics[width=\columnwidth,draft=false]{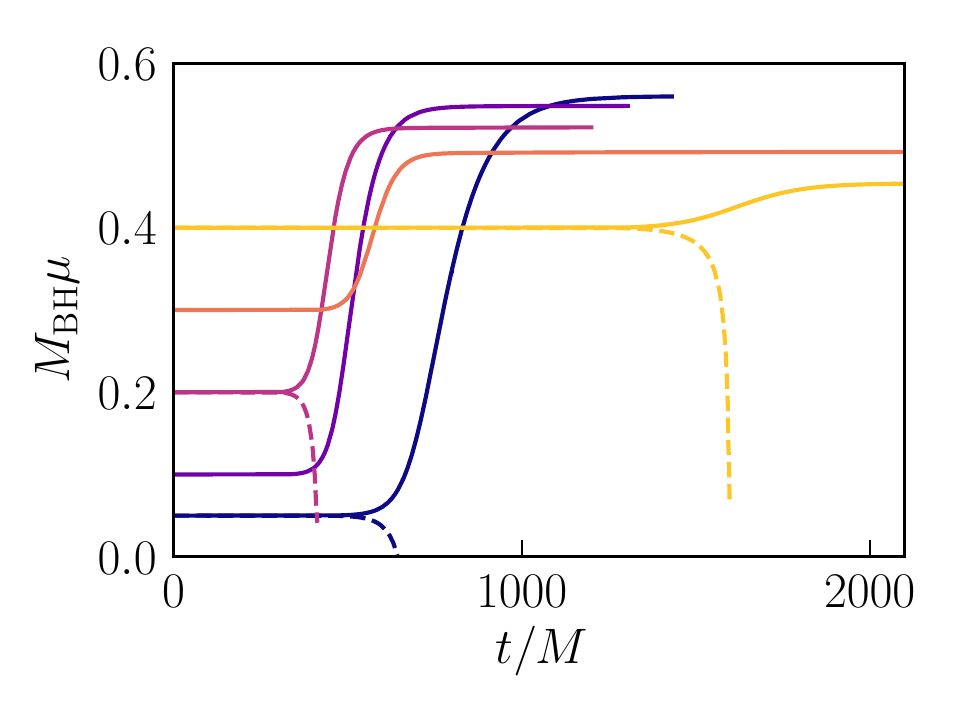}
\includegraphics[width=\columnwidth,draft=false]{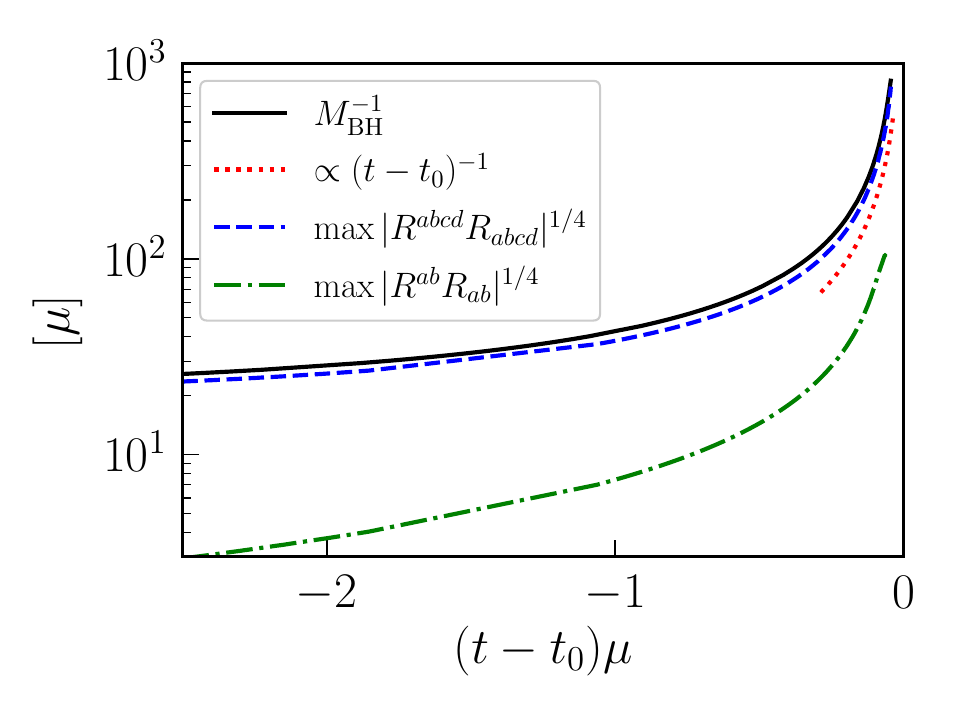}
\end{center}
\caption{
Top: the BH mass as a function of time during the nonlinear
    development of the spherically symmetric instability in Einstein-Weyl
    gravity for different values of $\alpha=\mu M$ from 0.05 to 0.4. Solid
    lines indicate unstable modes that cause the BH mass to grow,
    eventually saturating with clouds of the massive tensor field surrounding
    them. Corresponding dashed lines indicate opposite sign modes (with the
    same linear instability rate) that cause the BH mass to shrink to
    zero as far as the evolution can be carried out. Bottom: the
    late-time development of the instability where the BH shrinks for
    $\alpha=0.05$. The BH mass appears to approach zero at time $t_0$
    linearly in harmonic time, with the maximum value of $|R_{abcd}R^{abcd}|$
    outside the apparent horizon blowing up like $M_{\rm BH}^{-4}$.  The Ricci
    tensor squared $R_{ab}R^{ab}$ is subdominant, but its maximum magnitude
    also increases at a similar rate.
\label{fig:mbh_a0}
}
\end{figure}

First considering nonspinning BHs, in Fig.~\ref{fig:mbh_a0}, 
we illustrate the evolution of the BH mass as a result of the spherically
symmetric instability for different values of 
$\alpha$.\footnote{
Note, since we have defined $\alpha:=\mu M$ in terms of the global spacetime mass, it remains 
fixed even as the black hole mass changes. 
} 
When the BH mass
grows, the instability eventually saturates with the development of a massive tensor cloud
of effectively negative energy, and a BH mass exceeding
the value where the instability shutoffs for the isolated case: $M_{\rm BH}\gtrsim 0.44/\mu$. 
However, when the value of $\alpha$ is much smaller than this, 
the BH mass can significantly overshoot this threshold before saturating, 
e.g. by $\approx 30\%$ for $\alpha=0.05$. 

In contrast, when the BH mass decreases, we find no evidence for a
saturation of the instability, and the BH mass appears to approach zero
in finite time (though, of course, at finite numerical resolution we are not able
to track the decrease to arbitrarily small values).  In
the bottom panel of Fig.~\ref{fig:mbh_a0}, we show a case where (through the
use of high resolution) we track the BH as its mass decreases by a
factor of $\approx 40$, and find that it roughly follows $M_{\rm BH} \propto
(t-t_0)$ (in harmonic time) at late times. We expect the curvature at the
horizon to diverge as the BH becomes arbitrarily small
since $R_{abcd}R^{abcd}=0.75M_{\rm BH}^{-4}$ at the horizon of a Schwarzschild
BH. As shown in Fig.~\ref{fig:mbh_a0}, we indeed find that the curvature
outside the shrinking apparent horizon blows up in this way, suggesting that
the end state will be a naked curvature singularity. 

At the linear level, when $M_{\rm BH} \ll 1/\mu$, the instability rate is
independent of the BH mass, i.e., $\omega_I\sim \mu$, and thus the instability
timescale becomes long compared to the dynamical timescale of the BH.  If one
na\"ively assumes (in analogy to the superradiance instability of spin-0 and
spin-1 fields) that in the nonlinear regime of the instability the spacetime
migrates through a sequence of quasi-stationary Schwarzschild BHs of
adiabatically varying mass, then this finite-time diverging behavior is
expected purely from the linear analysis. Since when $M_{\rm BH} \ll 1/\mu$,
$\dot{M}_{\rm BH}=-\dot{M_c}=-C\mu M_c$ where $C$ is a numerical constant
independent of the BH mass $M_{\rm BH}$ and cloud mass $M_c$, in this adiabatic
approximation the BH's mass decreases to zero in finite time roughly as 
\begin{equation}
M_{\rm BH}\approx M_{\rm BH}^{t=0}+M_c^{t=0}(1-e^{C\mu t}) \ . 
\end{equation}
Expanding this expression around the time $t_0$ when $M_{\rm BH}=0$ gives
$M_{\rm BH} \propto (t-t_0)$, as found in the Einstein-Weyl evolutions.  This
means that generically a nonlinear theory must exhibit some strong backreaction
at small scales in order for black holes to avoid the fate found here.

\begin{figure}
\begin{center}
\includegraphics[width=\columnwidth,draft=false]{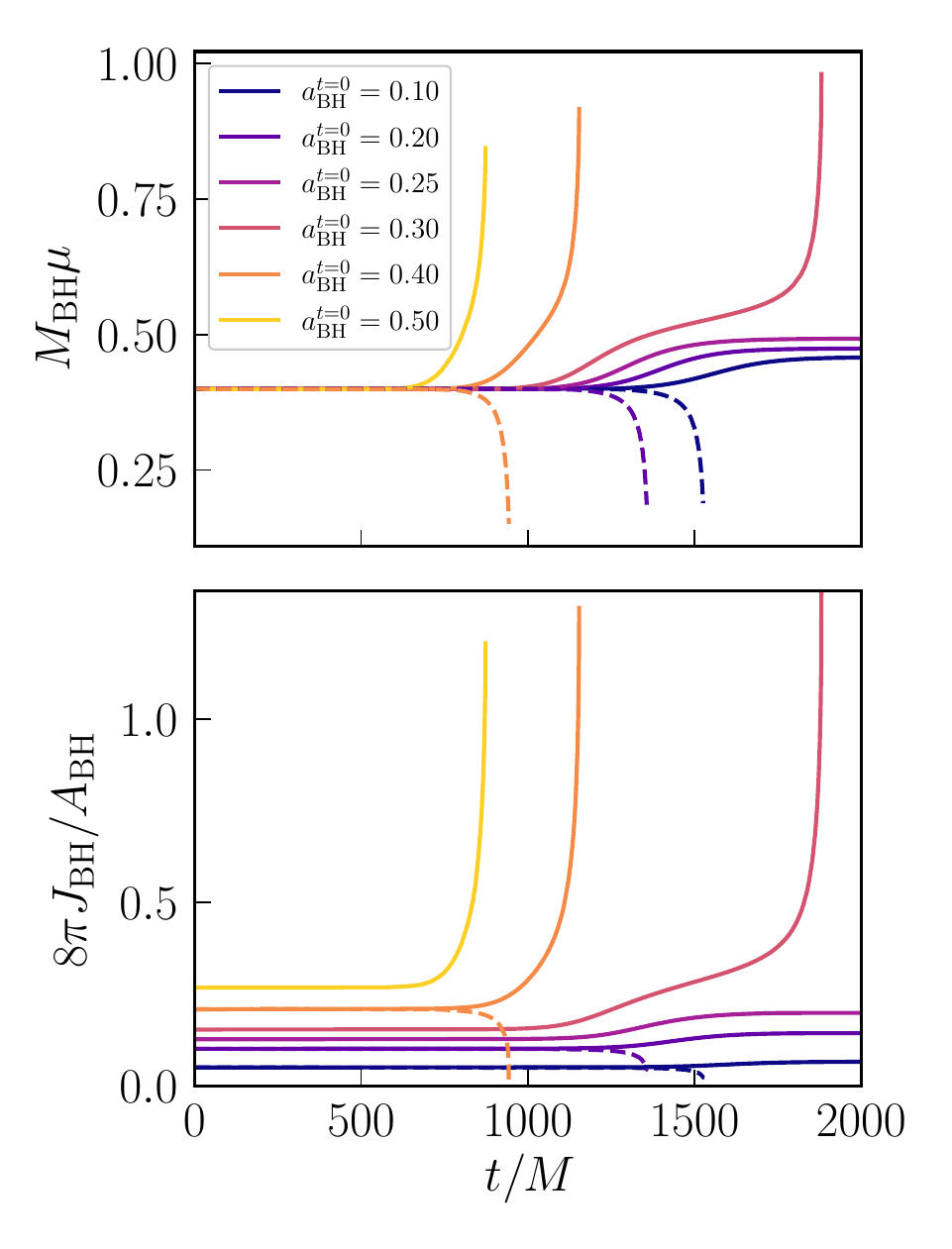}
\end{center}
\caption{
The nonlinear development of the axisymmetric ($m=0$) instability of BHs in
    Einstein-Weyl gravity for $\alpha=0.4$ and different initial values of the
    dimensionless BH spin $a_{\rm BH} \in[0.1, 0.5]$.  We
    show the evolution of the BH mass (top panel) and angular momentum
    normalized by the apparent horizon area (bottom panel), with the same color
    used in both panels for each case.  Solid lines indicate unstable modes
    that cause the BH mass and spin to grow. The lower spin cases saturate, as
    in the nonspinning cases. However, the higher spin cases exhibit rapidly
    increasing masses, and violate the extremality bound $J_{\rm BH}\leq A_{\rm
    BH}/(8\pi)$, and we are eventually unable to continue the evolution.
    Corresponding dashed lines indicate select cases with opposite sign modes
    that cause the BH spin and mass to decrease toward zero, in a similar
    manner to the nonspinning cases shown in Fig.~\ref{fig:mbh_a0}.
\label{fig:mbh_abh}
}
\end{figure}

Next, we consider the axisymmetric instability of spinning BHs.  For
simplicity, we fix $\alpha=0.4$ and vary the initial value of $a_{\rm BH}$. The
resulting evolution of the BH mass and angular momentum is shown in
Fig.~\ref{fig:mbh_abh}. For the sign of the initial perturbation where the BH
mass decreases, we find that the spin decreases as well (in this theory, the
massive tensor degree of freedom can carry away angular momentum even in
axisymmetry), rapidly approaching zero. Thus, we expect that these cases will
behave similarly to the case without angular momentum, approaching a zero-mass,
nonspinning BH in finite time. In contrast, for the opposite sign
perturbation, the BH mass and spin both increase. For smaller initial spins, we
again find that the solutions eventually saturate with a larger mass BH
surrounded by a massive tensor cloud. However, for larger initial spins, we
find that the mass and angular momentum rapidly increase, with the latter
reaching super-extremal values: $J_{\rm BH} > A_{\rm BH}/(8\pi)$ (note that, by
construction, $J_{\rm BH}\leq M_{\rm BH}^2$)~\cite{Booth:2007wu}.  In these cases, our evolutions
eventually breakdown (primarily due to not being able to track the apparent
horizon), and we leave the question of the ultimate fate of these cases to
future work. 

Here, we have restricted our spacetime to be axisymmetric ($m=0$), precluding the effect
of the superradiant instability. However, we do not expect this restriction to significantly
affect the cases we consider as the superradiant instability would operate on much longer timescales
than considered here. For the spinning cases we study, when the massive spin-2 cloud grows with positive
energy, it rapidly spins down the black hole, making superradiance irrelevant; when the massive
spin-2 cloud grows with negative energy, the black hole either rapidly reaches a super-extremal state where
we can no longer follow it, or it saturates at a moderate spin where (assuming the instability rates
are roughly comparable to the equivalent values in Kerr) superradiance would be slow.
 
\ssec{Discussion and Conclusion}%
Using time domain evolutions to study
the linear regime of theories propagating a massive tensor on the background of a spinning
BH, we find that the monopolar instability dominates over the superradiant instability
in much of the parameter space. 
Therefore, one cannot use the latter to place constraints
on massive spin-2 fields without taking into account
the backreaction of the former.
Note that, for $\alpha\ll 1$, the monopolar instability always dominates,
regardless of the BH spin, and has a timescale that is \textit{independent} of the BH's mass~\cite{Babichev:2013una,Brito:2013wya}, 
$\tau_{\rm mono.}\approx10 \ \text{s} \ (10^{-16}\ \text{eV}/{m_b})$.
As a result, all BHs, from solar-mass to supermassive ones with
$M_{\rm BH}=10^{10}M_\odot$, are unstable to the monopolar instability with
timescales smaller than the Salpeter accretion time: $\tau_{\rm mono.}<4.5
\times 10^7 \ \text{yrs}$, for a spin-2 field of mass 
$6\times 10^{-21}\ {\rm eV}>m_b>7\times 10^{-31}$ eV.  
For spin-2 masses heavier than this, we
find the superradiance instability becomes relevant in some parts of the
parameter space with rapidly spinning BHs and
$\alpha\gtrsim\mathcal{O}(1)$.  Finally, for $m_b\gtrsim 5\times 10^{-11}$
eV, even a light BH of mass $5\ M_\odot$ is stable to monopolar modes,
but may be unstable to superradiant modes with higher azimuthal numbers.  We
also find that higher azimuthal numbers are required for the superradiant
instability to be relevant at moderately high spins (e.g. $m\geq 3$ for $a_{\rm
BH}\lesssim 0.7$, as typical for the remnant of a low-spin, quasicircular
binary BH merger). Hence, an interesting direction for future
work is to determine the growth rates for these more precisely. 

As a simple model of the backreaction of the $m=0$ instability, we carry
out evolutions in Einstein-Weyl gravity. Though the backreaction will be
different in other (and in particular, ghost-free) nonlinear theories, there
are several features that we find that can already be anticipated from the
linear theory, and thus may be more generic.  Due to the short
timescales, we find that the BH can noticeably overshoot the values
where the linear analysis, considering the BH in isolation, would
indicate the instability shuts off. This contrasts with the spin-1 (and
presumably spin-0) superradiant instability, where the timescales are
longer~\cite{East:2017ovw,East:2018glu}. This overshooting happens when the BH 
mass grows to a larger value with a surrounding ghost spin-2 field cloud,
and the instability saturates outside the regime where
the linear analysis would indicate the superradiant instability is active.
On the other hand, when the spin-2 field grows at the expense of decreasing the BH mass
(as one would also expect for a nonghost field), this leads the BH to
approach zero mass in finite time.  This is consistent with the fact 
that, in the limit of small BH mass, the linear instability rate
has a nonzero value $\sim\mu$. Notably, the
Gregory-Laflamme instability of a black string~\cite{Gregory:1993vy}, which has
the same linear structure as the monopolar instability, also has a similar
fate: points along the black string shrink to zero radius in
finite time, also leading to a mild, zero-mass, naked curvature
singularity~\cite{Lehner:2010pn,Figueras:2022zkg}.
This type of singularity is also what arises in critical gravitational collapse~\cite{Choptuik:1992jv}.

For future work, it would be interesting to study the development of the
monopolar BH instability in other nonlinear theories, as well as to better
understand the fate of the BHs that we found to be spun up to extremal values by the
instability. One might also expect BHs to be spun up in theories where the instability decreases the BH mass
without extracting significant angular momentum.
It would also be interesting to follow the nonlinear development
of the superradiant instability in the regimes identified here where it
dominates. Of particular observational importance is the (likely
model-dependent) question of the gravitational wave signatures of these
different instabilities. A theory propagating both a massive and a massless
spin-2 field has, in general, more than two gravitational wave polarization
states with nontrivial dispersion relations
\cite{Bogdanos:2009tn,Tachinami:2021jnf}, which gravitational wave detectors
could, in principle, be sensitive to.

\acknowledgments
We thank Shinji Mukohyama, Claudia de Rham, Andrew Tolley, and Ramiro Cayuso
 for useful discussions. We thank Oscar Dias, Giuseppe Lingetti, Paolo Pani, and Jorge Santos
for sharing their numerical data with us.
We acknowledge support from an NSERC Discovery grant. Research at Perimeter Institute is
supported in part by the Government of Canada through the Department of
Innovation, Science and Economic Development Canada and by the Province of
Ontario through the Ministry of Colleges and Universities. This research was
undertaken thanks in part to funding from the Canada First Research Excellence
Fund through the Arthur B. McDonald Canadian Astroparticle Physics Research
Institute. This research was enabled in part by support provided by SciNet
(www.scinethpc.ca), Calcul Québec (www.calculquebec.ca), and the Digital
Research Alliance of Canada (www.alliancecan.ca). Simulations were performed on
the Symmetry cluster at Perimeter Institute, the Niagara cluster at the
University of Toronto, and the Narval cluster at the École de technologie
supérieure in Montreal.

\bibliographystyle{apsrev4-1.bst}
\bibliography{ref,bib}

\appendix
\section{Evolution Equations and Numerical Scheme} 
\label{app:eqns}
For the linear massive tensor field evolutions, we numerically evolve $H_{ab}$
and $\partial_t H_{ab}$ according to Eq.~\eqref{eqn:lin_eom} on the background of
a Kerr BH.
As in harmonic and Z4 formulations of general relativity~\cite{Gundlach:2005eh}, we find it necessary in BH spacetimes to add additional 
terms to Eq.~\eqref{eqn:lin_eom} that serve to damp violations of the constraint $\nabla_a H^{ab}=0$:
\begin{align}
    \Box H_{ab} =& \mu^2 H_{ab}-2R_{acbd}H^{cd} \nonumber \\ -&\kappa \left(n_a \nabla^c H_{bc} + n_b \nabla^c H_{ac}-\frac{1}{2}g_{ab}n^c\nabla^d H_{cd} \right ) \ .
    \label{eqn:lin_eom_cd}
\end{align}
The terms on the second line will converge to zero as $\nabla_a H^{ab} \rightarrow 0$, but serve to suppress constraint
violating modes that might otherwise grow, damping them on a timescale $1/\kappa$.
In this study, we typically set $\kappa \sim 1/M_{\rm BH}$.

In order to consider different unstable modes in isolation, and to reduce the
computational expense of the simulations, we consider cases where $H_{ab}$ has
an azimuthal symmetry with azimuthal number $m$.  Under this assumption,
following Ref.~\cite{East:2017mrj}, the computational domain can be reduced to
two spatial dimensions by introducing, for book keeping purposes, an imaginary
component to the massive tensor, $H_{ab}=H^R_{ab}+iH^I_{ab}$, and requiring
that  $\mathcal{L}_\phi H_{ab} = i m H_{ab}$, where $\mathcal{L}_\phi$ is the Lie
derivative with respect to the axisymmetric Killing vector. We use Cartesian
coordinates, but restrict our domain to the half plane with $0\leq x < \infty$,
$y=0$, and $-\infty < z < \infty$. By evolving both the real and imaginary part
of $H_{ab}$ according to Eq.~\eqref{eqn:lin_eom_cd}, we can compute out-of-plane
derivatives $\partial_y H_{ab}$ using the symmetry assumption. On the symmetry
axis, we apply regularity conditions to the components of $H_{ab}$. These expressions were obtained following the procedure in appendix A of Ref.~\cite{Pretorius:2004jg}.

For the nonlinear evolutions in Einstein-Weyl gravity, our evolution variables
are $\{g_{ab}, \partial_t g_{ab}, R_{ab}, \partial_t R_{ab}\}$. 
Following Ref.~\cite{Noakes:1983xd}, the metric
is evolved in the generalized harmonic formulation, where the gauge degrees are freedom
are fixed by requiring that $\Box x^a=H^a$, where $H^a$ are specified functions of the metric.
The only difference is that $R_{ab}$ now acts as a source term when evolving the metric:
\begin{align}
    g^{cd}\partial_c\partial_dg_{ab}=
-	2\nabla_{(a}H_{b)}
+	2H_c\Gamma^c_{ab}
-	2\Gamma^c_{da}\Gamma^d_{cb}
    \nonumber\\
-	\kappa\left(
            n_aC_b
    +	n_bC_a
    -	n_cC^cg_{ab}
    \right)
    \nonumber\\
-	2\partial_cg_{d(a}\partial_{b)}g^{cd}
      -2R_{ab}
    ,
    \label{eqn:gh_eom}
\end{align}
where we have also included the usual terms to damp the generalized harmonic constraint $C^a=H^a-\Box x^{a}$
~\cite{Gundlach:2005eh}.

The Ricci tensor is evolved according to Eq.~\eqref{eqn:nonlin_eom}, but with the addition of the same constraint
damping term as in Eq.~\eqref{eqn:lin_eom_cd}. The Riemann tensor term in Eq.~\eqref{eqn:nonlin_eom} is calculated
from the metric and its derivatives, but with the second time derivatives of $g_{ab}$ being determined
by substituting in Eq.~\eqref{eqn:gh_eom}.
Recall that, under our assumptions, $R_a^a=0$. We evolve all ten components of $R_{ab}$, but subtract
out any trace component (due to truncation error) at each time step.
See Refs.~\cite{Held:2021pht,Held:2023aap} for related evolution schemes.

For the linear evolutions, we use either harmonic coordinates~\cite{Cook:1997qc} or (when the BH spin is large) Cartesian Kerr-Schild coordinates~\cite{1965cngg.conf..222K}.
For the nonlinear evolutions, we begin with a BH in harmonic coordinates and use the harmonic gauge: $\Box x^a=0$.
Spatial derivatives are discretized using fourth-order finite difference stencils, while time stepping is performed using fourth-order
Runge Kutta. We use adaptive mesh refinement to concentrate numerical resolution around the BH, but the interpolation
for the refinement boundaries is only accurate to third-order in the size of the time step.

For the nonlinear evolutions in axisymmetry, we track the BH apparent horizon by finding the outermost marginally outer trapped surface using a flow method~\cite{Pretorius:2004jg}. Integrating over the horizon surface, we compute the area $A_{\rm BH}$, and the angular momentum 
\begin{equation}
J_{\rm BH} = \frac{1}{8 \pi} \int \hat\phi_i K^{ij} dA_j \ ,
\end{equation}
where $K^{ij}$ is the extrinsic curvature and $\hat \phi^i$ is the axisymmetric Killing vector. From these quantities we can define a BH mass through the Christodoulou formula
\begin{equation}
    M_{\rm BH} := \left (\frac{A_{\rm BH}}{16\pi}+\frac{4 \pi J_{\rm BH}^2}{A_{\rm BH}} \right)^{1/2} \ .
\end{equation}
Note that, with this definition, the dimensionless spin is given by
\begin{equation}
    a_{\rm BH}=\frac{J_{\rm BH}}{M_{\rm BH}^2}=\frac{2j}{1+j^2}\  {\rm with} \ j=\frac{8\pi J_{\rm BH}}{A_{\rm BH}}. 
\end{equation}
Hence, $a_{\rm BH}$ obtains a maximum of $a_{\rm BH}=1$ at $j=1$ and decreases for larger $j$.

\section{Numerical Convergence} 
\label{app:conv}

For most of the evolutions we perform, we use a computational domain with seven or eight levels of mesh refinement centered on the BH,
and where the grid spacing on the finest level is between $dx\approx 0.04 M$ and $0.02M$.  
For nonlinear evolutions where the BH shrinks, we add additional resolution and mesh refinement levels. In the most extreme case, shown in the bottom panel of Fig.~\ref{fig:mbh_a0}, we have $dx \approx 5\times 10^{-4} M$ on the finest level used.
We also perform resolution studies of select cases to check for convergence and to estimate numerical errors.

\begin{figure}
\includegraphics[width=\columnwidth,draft=false]{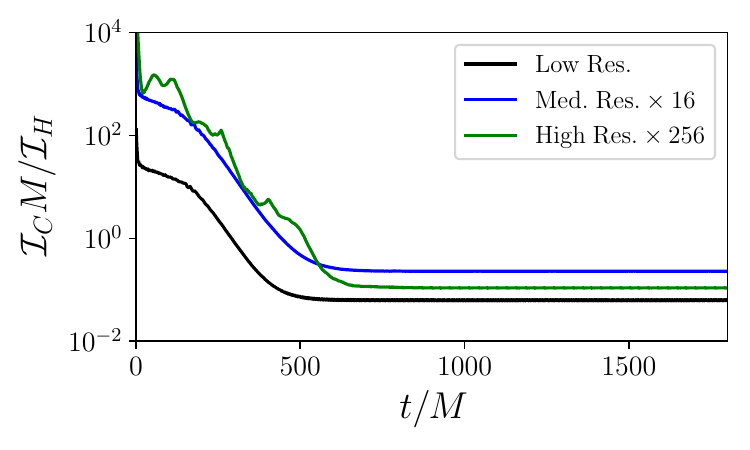}
\caption{The evolution of the Bianchi constraints $\mathcal{I}_C:=\int d^3x
[\sum_a(\nabla_b H^{ab}_R)^2]^{1/2}+(R\leftrightarrow I)$, normalized by the
field amplitude $\mathcal{I}_H:=|\int d^3x
H_{ab}^RH^{ab}_R|^{1/2}+(R\leftrightarrow I)$, throughout the development of
the $m=1$ superradiance instability of a massive spin-2 field on a Kerr BH
background with spin $a_{\rm BH}=0.95$ and $\alpha=0.7$. Since the initial
perturbation used to trigger the linear instability does not satisfy the
constraints, they do not converge at early times. At later times, both $\mathcal{I}_C$
and $\mathcal{I}_H$ grow exponentially at the instability rate. 
The different resolutions have been scaled by a factor consistent with fourth
order convergence.}
\label{fig:lin_cnst_conv}
\end{figure}

\begin{figure}
\begin{center}
\includegraphics[width=\columnwidth,draft=false]{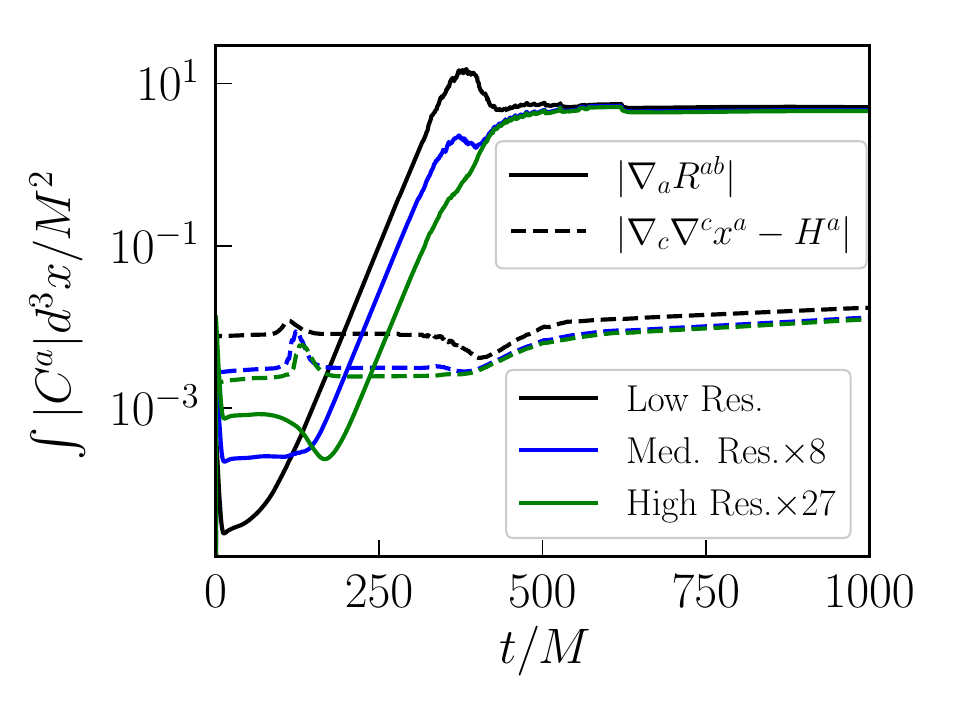}
\end{center}
\caption{
The convergence of the constraints with increasing resolution for a nonlinear evolution in Einstein-Weyl gravity
of a nonspinning BH with $\alpha=0.2$. In this case (corresponding to the solid magenta curve in Fig.~\ref{fig:mbh_a0}),
the BH mass increases and saturates at a larger value. We show both the norm of the Bianchi constraint $\nabla_a R^{ab}=0$ (solid lines) and the generalized harmonic constraint $H^a-\Box x^{a}=0$ (dashed lines) integrated on the domain. The constraints converge at between third and fourth order, with the different resolutions scaled by a factor consistent with the former. Due to the fact that the initial perturbation we use does not satisfy the constraints, there is nonconvergence in the Bianchi constraint at early times, but this is exponentially
small compared to the subsequent truncation error, and does not affect the convergence at later times. 
\label{fig:cnst_conv}
}
\end{figure}

In Fig.~\ref{fig:lin_cnst_conv}, we show a norm of the Bianchi constraint
violation $\sum_b |\nabla_a H^{ab}|/4$, integrated over the domain, during a
linear evolution case with $m=1$ (summing the contributions from the
real and imaginary components).  We normalize this quantity by the norm of
the massive tensor $|H^{ab}H_{ab}|$, as the whole solution is exponentially
increasing due to the instability.  As evident from the figure, the relative
constraint violation at later times is roughly constant in time, and converging
to zero with increasing resolution.  We show a nonlinear evolution with 
Einstein-Weyl gravity in Fig.~\ref{fig:cnst_conv}. There, in addition to the
Bianchi constraint violation, we show the norm of the generalized harmonic
constraint. As can be seen from the plot, the constraints converge to zero with
increasing resolution at the expected rate (between third and fourth order).

For several of the linear evolutions cases, we extract the instability growth
rate at three different resolutions, and use Richardson extrapolation to
estimate the truncation error in these quantity. The results are shown in
Table~\ref{tab:table1}. The errors in the instability rates for the default
resolutions do vary noticeably with $\alpha$ for the considered values, ranging from order $10\%$ to a
few percent or smaller, but these provide a rough estimate for the errors
across the parameter space. Generally, we find that for larger BH spin and
$\alpha$, the uncertainties are larger. 
\begin{table}[h!]
  \centering
  \caption{A comparison of the instability growth rate measured at the default resolution,
          to the Richardson extrapolated value using three different resolutions.  This gives a measure of the truncation error, shown as a percent in the last column.}
  \label{tab:table1}
  \begin{tabular}{c|c|c|c|c|c}
    \hline
                &              &     & $M\omega_I$: default & Richardson & Error\\
       $\alpha$ & $a_{\rm BH}$ & $m$ & resolution & extrapolated version & (\%) \\
    \hline
      0.4 & 0.99 & 0 & $7.1872\times10^{-2}$ & $7.1866\times10^{-2}$ & 0.01 \\
      1.6 & 0.998 & 0 & $7.65\times10^{-3}$ & $8.07\times10^{-3}$ & 5.2  \\
      0.7 & 0.95 & 1 & $1.27\times 10^{-2}$ & $1.25\times 10^{-2}$ & 1.6 \\
      0.9 & 0.998 & 1 & $1.91\times 10^{-2}$ & $2.01\times 10^{-2}$ & 5.0 \\
      1.0 & 0.95 & 2 & $8.80\times 10^{-4}$ & $8.18\times 10^{-4}$ & 8.0 \\ 
      1.5 & 0.99 & 2 & $1.59\times 10^{-3}$ & $1.40\times 10^{-3}$ & 14 \\
      1.4 & 0.998 & 2 & $2.19\times 10^{-3}$ & $2.13\times 10^{-3}$ & 3.0 \\
      1.9 & 0.99 & 3 & $1.19\times 10^{-4}$ & $1.08\times 10^{-4}$ & 10 \\ \hline
  \end{tabular}
\end{table}

\section{Real frequencies and comparison to literature} \label{app:modes}

\begin{figure}[t]
\includegraphics[width=\columnwidth,draft=false]{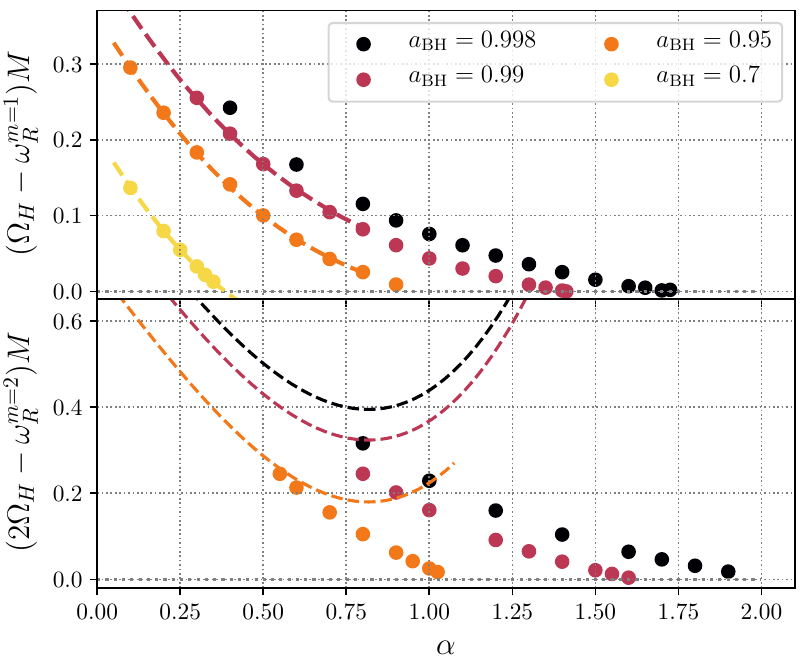}
    \caption{The real part of the frequency $\omega_R$ of the most unstable
    $m=1$ (\textit{top}) and $m=2$ (\textit{bottom}) superradiant modes of a
    massive spin-2 field around Kerr BHs of various dimensionless spins
    $a_{\rm BH}$. Circular points correspond to our time-domain results. 
    The dashed lines are interpolations of the numerical results in Ref.~\cite{Dias:2023ynv}, for the
    most unstable $m=1$ mode, and analytic expressions determined in
    Ref.~\cite{Brito:2020lup}, for the most unstable $m=2$ superradiant state.}
\label{fig:real_freq}
\end{figure}

\begin{figure}[t]
\includegraphics[width=\columnwidth,draft=false]{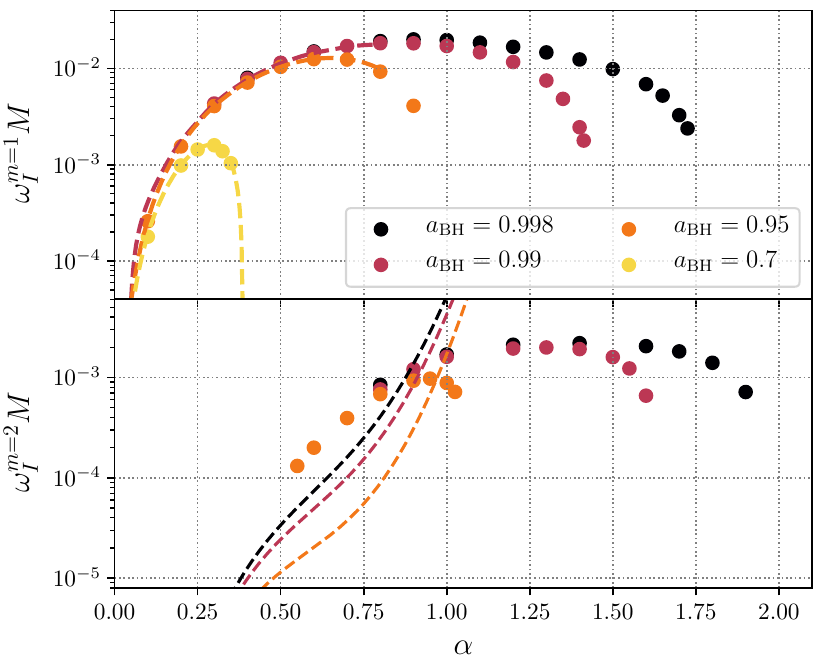}
    \caption{
    The top and bottom panels correspond to the imaginary part of the frequency $\omega_I$ of the
    $m=1$ and $2$ modes, respectively, shown in Fig.~\ref{fig:real_freq} (circular points). 
    For comparison, we show the interpolated numerical data from Ref.~\cite{Dias:2023ynv} (top)
    and the analytic expressions from Ref.~\cite{Brito:2020lup} (bottom) for these unstable modes (dashed lines).
    }
\label{fig:imag_freq}
\end{figure}

For completeness, we provide the real parts of the frequencies associated with
the fastest growing $m=1$ and $m=2$ superradiant modes (the $m=0$ modes
have zero real frequency) here, and compare these,
as well as the growth rates, to results obtained in
Refs.~\cite{Dias:2023ynv,Brito:2020lup}. 
To that end, we show the real parts of
the frequencies in Fig.~\ref{fig:real_freq} and the imaginary parts in
Fig.~\ref{fig:imag_freq}.

In Ref.~\cite{Dias:2023ynv}, the frequencies, $\omega_R$ and $\omega_I$, of the
most unstable $m=1$ superradiant modes were determined by solving the elliptic
equations governing the eigenvalue problem for $\alpha\leq 0.8$ and $a_{\rm BH}\leq 0.99$.
Comparing their results (interpolated to some of the specific values of $a_{\rm BH}$ we consider) in the region where they overlap with ours 
 in the top panels
of Figs.~\ref{fig:real_freq} and~\ref{fig:imag_freq}, we find 
good agreement (to within roughly $5\%$, and consistent with the expected truncation error).

In Ref.~\cite{Brito:2020lup}, analytic estimates for
$\omega_R$ and $\omega_I$ of the most unstable $m=2$ superradiant modes were
obtained in the $\alpha\ll 1$ limit. Up to $\alpha\approx 0.5$, these
expressions for $\omega_R$ match our results (compare, in particular, the $m=2$
and $a_{\rm BH}=0.95$ family of modes in Fig.~\ref{fig:real_freq}). The
estimates for the growth rates of these configurations, on the other hand, are
inconsistent with our time-domain predictions at $\alpha\approx 0.5$; we expect
that the agreement would be better in the $\alpha<0.5$ region of the parameter
space. Our data for the $m=3$ superradiant mode covers $\alpha\geq 1.3$, a
regime in which we do not find good agreement with the perturbative estimates
valid in the $\alpha\ll 1$ limit.

Finally, we mention that the numerical values of $\omega_R$ and $\omega_I$ calculated in this work for $m=0$, 1, 2, and 3
(as well as some lower values of $a_{\rm BH}$ for $m=0$ omitted here for clarity) are available at Ref.~\cite{spin2_data}.

\end{document}